**Multi-IMPT: a biologically equivalent approach to proton ARC therapy**

Nimita Shinde, Yanan Zhu, Wei Wang, Wangyao Li, Yuting Lin, Gregory N Gan, Christopher Lominska, Ronny Rotondo, Ronald C Chen, and Hao Gao[*]

Department of Radiation Oncology, University of Kansas Medical Center, USA

* Corresponding author:

Hao Gao, Department of Radiation Oncology, University of Kansas Medical Center, USA.

Email address: hgao2@kumc.edu

**Conflict of Interest Statement**

None.

**Ethical Statement:** This research was carried out under Human Subject Assurance Number 00003411 for University of Kansas in accordance with the principles embodied in the Declaration of Helsinki and in accordance with local statutory requirements. Consent was given for publication by the participants of this study.

**Acknowledgment**

The authors are very thankful for the valuable comments from reviewers. This research is partially supported by the NIH grants No. R37CA250921, R01CA261964, and a KUCC physicist-scientist recruiting grant.

**Abstract.**

**Objective:** Proton spot-scanning arc therapy (ARC) is an emerging modality that can improve the high-dose conformity to targets compared with standard intensity-modulated proton therapy (IMPT). However, the efficient treatment delivery of ARC is challenging due to the required frequent energy changes during the continuous gantry rotation. This work proposes a novel method that delivers a multiple IMPT (multi-IMPT) plan that is equivalent to ARC in terms of biologically effective dose (BED).

**Approach:** The proposed multi-IMPT method utilizes a different subset of limited number of beam angles in each fraction for dose delivery. Due to the different dose delivered to organs at risk (OAR) in each fraction, we optimize biologically effective dose (BED) for OAR and the physical dose delivered for target in each fraction. The BED-based multi-IMPT inverse optimization problem is solved via the iterative convex relaxation method and the alternating direction method of multipliers. The effectiveness of the proposed multi-IMPT method is evaluated in terms of dose objectives in comparison with ARC.

**Main results:** Multi-IMPT provided similar plan quality with ARC. For example, multi-IMPT provided better OAR sparing and slightly better target dose coverage for the prostate case; similar dose distribution for the lung case; slightly worse dose coverage for the brain case; better dose coverage but slightly higher BED in OAR for the head-and-neck case.

**Significance:** We have proposed a multi-IMPT approach to deliver ARC-equivalent plan quality.

# 1. Introduction

Sandison et al. [1] proposed passive-scattering-based proton arc radiotherapy (PS-ARC) and demonstrated an approach could improve the target dose conformity. However, the implementation of PS-ARC [2, 3, 4] faced several limitations including the need to change beam compensator and range modulation wheel during gantry rotation.

The technological challenges posed by PS-ARC were addressed by proton spot-scanning arc therapy (ARC) [5, 6, 7, 8, 9, 10] that uses modern scanning nozzles and does not require the compensator and range modulation wheel. ARC emerges as an advanced treatment method that can provide conformal dose distribution and spare organs at risk (OAR) adjacent to targets [5, 6, 8, 11, 12, 13]. However, during the ARC delivery, while the gantry rotates continuously, the energy changes can happen frequently, which can substantially impact the delivery efficiency due to energy switching time, e.g., each energy switching-up of around 5s and each energy switching-down time of around 0.5s.

Various energy layer optimization methods have been proposed to minimize the number of energy switching-ups and energy layer changes to improve ARC delivery efficiency [1, 6, 14, 15, 16, 17], including mono-energy-per-control-point regularization [6], heuristic algorithms [16], and mixed-integer programming approach [17]. However, maintaining plan quality while simultaneously reducing the number of energy changes remains challenging.

In this work, we propose a novel method called multi-IMPT that is biologically equivalent to ARC in terms of plan quality. The multi-IMPT consists of multiple IMPT plans alternatingly delivered over the entire course, with each IMPT plan consisting of a different subset of limited number of field angles. The key difference between multi-IMPT and ARC is the dose to normal tissues and OAR, resulting in the need to model the effect of temporal fractionation for multi-IMPT. Thus, biologically effective dose (BED), calculated using linear and quadratic (LQ) model for normal tissues, will be used to demonstrate the biological equivalence of multi-IMPT and ARC.

## 2. Problem Formulation

In this section, we start by defining the parameters for the optimization problem (including the definition of the BED), followed by defining the constraints to minimize the BED in OAR, and finally introducing the complete optimization model for multi-IMPT.

*2.1. Defining parameters, decision variables and constraints in the multi-IMPT optimization problem*

We now define all the quantities associated with our optimization problem followed by defining the constraints in the problem.

Parameters and decision variable:

1. $M = \{1, \ldots, M\}$: set of indices of OAR
2. For $m \in M$, $n_m$: number of voxels in $m$-th OAR
3. $A^{tm} \in R^{n_m \times k_t}$: dose influence matrix of $m$-th OAR during fraction $t$; $k_t$ is the number of beams in the active fields in fraction $t$; $A_j^m$: $j$-th row of the matrix $A^m$ and corresponds to the $j$-th voxel in OAR $m$.
4. $A^{t0} \in R^{n_0 \times k_t}$: dose influence matrix corresponding to tumor/target during fraction $t$; $n_0$ is the number of voxels in the tumor.
5. $T$: number of fractions
6. (Decision variable) $u_t \in R^{k_t}$: spot intensity vector in fraction $t$, for $t = 1, \ldots, T$

Biologically Effective Dose (BED) [18, 19, 20] and physical dose (d):

1. **BED in OAR:** For OAR $m$, let $\alpha_m, \beta_m$ be the parameters of the well-known LQ-model that is used to define BED. Define $\rho_m = 1/(\alpha_m/\beta_m)$. Under the LQ model, the total biologically effective dose (BED) delivered to the $j$-th voxel in OAR $m$ is $\sum_{t=1}^{T}(A_j^{tm}u_t + \rho_m(A_j^{tm}u_t)^2)$. During the experiments, we set $\alpha_m/\beta_m$ value to 2 Gy for all $m$.
2. **Physical dose delivered to target:** The physical dose delivered to each target voxel $j \in [n_0]$ in each fraction is calculated as $d_j^{t0} = A_j^{t0}u_t$.

Constraints in the model:

1. **BED-max constraint for OAR [21, 22, 23, 24]:** Let $M_1$ be the set of OAR that are highly sensitive to radiation, and their function is hampered even when a single voxel is damaged by radiation. For such OAR, the BED-max constraint bounds the maximum BED ($BED_{max}^m$) delivered to each voxel in OAR $m$. Thus, we define

$$\sum_{t=1}^{T} A_j^{tm} u_t + \sum_{t=1}^{T} \rho_m (A_j^{tm} u_t)^2 \leq BED_{max}^m \quad \forall\, m \in M_1, j \in [n_m].$$

2. **BED-mean constraint for OAR** [21, 22, 23, 24]: Let $M_2$ be the set of OAR whose small portion can be damaged without affecting their function. For such OAR, the BED-mean constraint bounds the mean BED ($BED_{mean}^m$) delivered to all voxels in OAR $m$. Thus, we define

$$\sum_{j=1}^{n_m}\sum_{t=1}^{T} A_j^{tm} u_t + \sum_{j=1}^{n_m}\sum_{t=1}^{T} \rho_m (A_j^{tm} u_t)^2 \leq n_m \times BED_{mean}^m \quad \forall\, m \in M_2.$$

3. **BED-DVH max constraint for OAR** [21, 22, 23, 24]: Consider the set of OAR $M_3$. The BED-DVH constraints states that for any OAR $m \in M_3$, at most $p$ fraction of voxels should receive BED larger than $BED_{dv}^m$, i.e., $BED_j^m = \sum_{t=1}^{T} A_j^{tm} u_t + \sum_{t=1}^{T} \rho_m (A_j^{tm} u_t)^2 \geq BED_{dv}^m$ for at most $p \times n_m$ voxels. One of the commonly used techniques to define the DVH max constraint is to first define the set of indices (called active index set) of voxels that violate the constraint. More precisely, let $[n_m']$ be the set of indices of voxels that are sorted in descending order of the BED delivered to the voxels in OAR $m$. The active index set is then defined as

$$\Omega^m = \{ j \in [n_m'] \mid j \geq p \times n_m, BED_j^m \geq BED_{dv}^m \}.$$

If the active index set, $\Omega^m$, is non-empty, the BED-DVH max constraint is be defined as

$$\sum_{t=1}^{T} A_j^{tm} u_t + \sum_{t=1}^{T} \rho_m (A_j^{tm} u_t)^2 \leq BED_{dv}^m \quad \forall\, m \in M_3, j \in \Omega^m.$$

4. **DVH min constraint for target** [25, 26]: DVH min constraint ensures that at least $p$ fraction of the target voxels receive physical dose larger than $d_{dv}^0$ in each fraction $t$, i.e., $d_j^{t0} \geq d_{dv}^0$, for at least $p \times n_0$ voxels. To define the DVH min constraint, we first define the active index set for the target as

$$\Omega^0 = \{ j \in [n_0'] \mid j \leq p \times n_m, d_j^{t0} \leq d_{dv}^0 \},$$

where $[n_0']$ is the set of indices of the target voxels sorted in the descending order of the dose delivered. The DVH min constraint is then defined as

$$d_j^{t0} \geq d_{dv}^0 \ \forall j \in \Omega^0.$$

5. **Max dose for target:** Finally, we define a constraint that bounds the maximum physical dose that can be tolerated by target dose in each fraction. In our model, we add the constraint

$$d_j^{t0} = A_j^{t0} u_t \leq 1.1 px \ \ \forall j \in [n_0], t \in [T],$$

i.e., the dose delivered to each target voxel $j$ in each fraction $t$ should not exceed 1.1 times the prescribed physical dose ($px$).

*2.2 Optimization problem*

Combining the constraints defined above, we now define the multi-IMPT optimization problem:

$$\min_{u_t} \sum_{t=1}^{T} ||A^{t0} u_t - px||_2^2$$

$$\text{s.t.} \sum_{t=1}^{T} A_j^{tm} u_t + \sum_{t=1}^{T} \rho_m (A_j^{tm} u_t)^2 \leq BED_{max}^m \ \forall m \in M_1, j \in [n_m],$$

$$\sum_{j=1}^{n_m} \sum_{t=1}^{T} A_j^{tm} u_t + \sum_{j=1}^{n_m} \sum_{t=1}^{T} \rho_m (A_j^{tm} u_t)^2 \leq n_m \times BED_{mean}^m \ \forall m \in M_2, \quad (1)$$

$$\sum_{t=1}^{T} A_j^{tm} u_t + \sum_{t=1}^{T} \rho_m (A_j^{tm} u_t)^2 \leq BED_{dv}^m \ \forall m \in M_3, j \in \Omega^m,$$

$$\sum_{t=1}^{T} A^{t0} u_t + \sum_{t=1}^{T} \rho_0 (A^{t0} u_t)^2 \geq BED_{dv}^0 \ \forall j \in \Omega^0,$$

$$A_j^{t0} u_t \leq 1.1 \quad x \ \ \forall j \in [n_0], t \in [T],$$

$$u_t \in \{0\} \cup [g, +\infty) \ \forall t = 1, \dots, T.$$

The last constraint in Eq. (1) defines a minimum-monitor-unit (MMU) constraint [27, 28, 29, 30, 31, 32] for $u_t$ with $g$ as the MMU threshold to ensure plan deliverability. The multi-IMPT model (Eq. (1)) is non-convex with quadratic constraints. Before we provide the solution methodology, we first provide a comparison of Eq. (1) with ARC.

**Comparison with ARC:** Note that, our decision variable (spot intensity vector $u_t$) is not the same in every fraction since we do not use the same fields in every fraction. Instead, we choose a different subset of limited number of beam angles in each fraction. Thus, multi-IMPT (Eq. (1)) differs from ARC, where all fields are active in each fraction resulting in equal dose per fraction, i.e., $u_t = u \ \forall t$. We should note that the size of the decision variable in the ARC model is much larger since it consists of spot intensities from each beam spaced at 15° interval over a 360° rotation resulting in one large optimization problem that needs to be solved to find optimal $u$. In contrast, in Eq. (1), the spot intensities in each fraction are defined from a small subset of beams. Thus, the size of the decision variables $u_t$ in Eq. (1) is much smaller than the size of decision variable $u$ in ARC model. Furthermore, as we see in Section 2.3, the optimization problem in Eq. (1) can be separated in $t$, resulting in multiple computationally cheaper and smaller optimization problems defined for each $t$.

*2.3 Solution algorithm*

To solve Eq. (1), we first introduce additional variables. Define $z_j^{tm} = A_j^{tm} u_t$ for all $j \in [n_m], m \in M_1 \cup M_2 \cup M_3, t \in [T]$, and $z_j^{t0} = A_j^{t0} u_t$ for all $j \in [n_0], t \in [T]$ and re-write Eq. (1) as

$$\min_{u_t, z_j^{tm}, z_j^{t0}} \sum_{t=1}^{T} ||A^{t0} u_t - px||_2^2$$

$$\text{s.t.} \sum_{t=1}^{T} (z_j^{tm}) + \sum_{t=1}^{T} \rho_m (z_j^{tm})^2 \leq BED_{max}^m \ \forall m \in M_1, j \in [n_m],$$

$$\sum_{j=1}^{n_m} \sum_{t=1}^{T} (z_j^{tm}) + \sum_{j=1}^{n_m} \sum_{t=1}^{T} \rho_m (z_j^{tm})^2 \leq n_m \times BED_{mean}^m \ \forall m \in M_2,$$

$$\sum_{t=1}^{T} (z_j^{tm}) + \sum_{t=1}^{T} \rho_m (z_j^{tm})^2 \leq BED_{dv}^m \ \forall m \in M_3, j \in \Omega^m, \qquad (2)$$

$$\sum_{t=1}^{T} (z_j^{t0}) + \sum_{t=1}^{T} \rho_0 (z_j^{t0})^2 \geq BED_{dv}^0 \ \forall j \in \Omega^0,$$

$$z_j^{t0} \leq 1.1 px \ \forall j \in [n_0], t \in [T],$$

$$z_j^{tm} = A_j^{tm} u_t \ \forall j \in [n_m], m \in M_1 \cup M_2 \cup M_3, t \in [T],$$

$$z_j^{t0} = A_j^{t0} u_t \ \forall j \in [n_0], t \in [T],$$

$$u_t \in \{0\} \cup [g, +\infty\} \ \forall t = 1, \dots, T.$$

We now solve Eq. (2) via iterative convex relaxation (ICR) method [33, 34] and alternating direction method of multipliers (ADMM) method [35, 36]. The method involves iteratively updating the active index sets (defined in Section 2.1) followed by sequentially updating each decision variable in the problem. To do so, we first define the augmented Lagrangian as

$$\min \frac{w_0}{n_0} \sum_{t=1}^{T} ||A^{t0} u_t - px||_2^2 + \frac{\mu_1}{2} \sum_{t=1}^{T} \sum_{m \in M_1 \cup M_2 \cup M_3} \frac{w_m}{n_m} ||A^{tm} u_t - z^{tm} + \lambda^{tm}||_2^2$$

$$+ \frac{\mu_2}{2} \sum_{t=1}^{T} \frac{w_0^1}{n_0} ||A^{t0} u_t - z^{t0} + \lambda^{t0}||_2^2 + \frac{\mu_3}{2} \sum_{t=1}^{T} \frac{w_0^2}{n_0} ||A^{t0} u_t - 1.1px + \gamma^t||_2^2 + \frac{\mu_4}{2} \sum_{t=1}^{T} ||u_t - y_t + \zeta_t||_2^2$$

$$\text{s.t.} \sum_{t=1}^{T} (z_j^{tm}) + \sum_{t=1}^{T} \rho_m (z_j^{tm})^2 \leq BED_{max}^m \ \forall\, m \in M_1, j \in [n_m],$$

$$\sum_{j=1}^{n_m} \sum_{t=1}^{T} (z_j^{tm}) + \sum_{j=1}^{n_m} \sum_{t=1}^{T} \rho_m (z_j^{tm})^2 \leq n_m \times BED_{mean}^m \ \forall\, m \in M_2, \quad (3)$$

$$\sum_{t=1}^{T} (z_j^{tm}) + \sum_{t=1}^{T} \rho_m (z_j^{tm})^2 \leq BED_{dv}^m \ \forall\, m \in M_3, j \in \Omega^m,$$

$$\sum_{t=1}^{T} (z_j^{t0}) + \sum_{t=1}^{T} \rho_0 (z_j^{t0})^2 \geq BED_{dv}^0 \ \forall\, j \in \Omega^0,$$

$$y_t \in \{0\} \cup [g, +\infty] \ \forall\, t = 1, \ldots, T.$$

In Eq. (3), $u_t, z^{tm}, z^{t0}, y_t$ are primal variables and $\lambda^{tm}, \lambda^{t0}, \gamma^t, \zeta_t$ are dual variables. Algorithm 1 provides a brief outline of the optimization method that solves Eq. (3). We explain each step in the appendix.

---

**Algorithm 1: Optimization method for solving Eq. (3)**

1. **Input:** Choose parameters $\mu_1, \ldots, \mu_4, w_0, w_m, w_0^1, w_0^2$
2. Initialization: Randomly initialize $u_t$. Choose iteration number $K$
3. Set $\lambda^{tm} = z^{tm} = A^{tm} u_t$, $\lambda^{t0} = z^{t0} = A^{t0} u_t$, $\zeta_t = y_t = u_t$, $\gamma_t = 1.1px$ for all $t$
4. For $k = 1, \ldots, K$
    a. Find active index sets $\Omega^m, \Omega^0$ for BED-DVH and DVH constraints as described in Section 2.1
    b. Update primal variables $u_t, z^{tm}, z^{t0}, y_t$ $\forall t$ by fixing all variables except one and solving the resulting minimization problem
    c. Update dual variables as follows:
    $$\lambda^{tm} = \lambda^{tm} + A^{tm} u_t - z^{tm}$$
    $$\lambda^{t0} = \lambda^{t0} + A^{t0} u_t - z^{t0}$$
    $$\gamma^t = \gamma^t + A^{t0} u_t - 1.1px$$
    $$\zeta_t = \zeta_t + u_t - y_t$$
5. **Output:** $u_t$

*2.4 Materials*

We show the equivalency of multi-IMPT and ARC in terms of the biologically effective dose for four clinical cases. For ARC, the control points for each beam are spaced at 15º intervals over a 360º rotation. For multi-IMPT, six beam angle combinations are used: (0º, 90º, 180º, 270º), (15º, 105º, 195º, 285º), (30º, 120º, 210º, 300º), (45º, 135º, 225º, 315º), (60º, 150º, 240º, 330º), and (75º, 165º, 255º, 345º), generating six different IMPT plans. We generate the ARC plan and multi-IMPT plans by solving Eq. (1) using the method described in Algorithm 1. The dose influence matrix is generated using MatRad [37], with spot width of 5 mm on 3 mm³ dose grid.

We consider four clinical cases with prescription dose and number of fractions given as: (1) prostate case (1.8 Gy x 25 fractions), (2) lung case (2 Gy x 30 fractions), (3) brain case (1.2 Gy x 60 fractions), (4) head and neck (HN) case (2 Gy x 35 fractions). The upper bound for each constraint on the BED delivered to the OAR are stated in Table 1-4 for the respective test cases. $BED_p$ denotes that at most $p$% of OAR voxels should receive BED greater than the value defined as the upper bound. For a fair comparison, we normalize all multi-IMPT and ARC plans so that 95% of the target volume receives 100% of the prescription dose. To quantify the plan quality, we compare the following quantities for the two plans: (a) Conformity Index (CI), (b) maximum dose delivered to tumor ($D_{max}$), (c) mean and max BED to OAR. CI is defined as $V_{100}^2/(V \times V'_{100})$, where $V_{100}$ is the target volume that receives at least 100% of the prescription dose, $V$ is the target volume, and $V'_{100}$ is the total volume that receives at least 100% of the prescription dose.

## 3. Results

**Prostate:** Table 1 presents the results of the comparison between the multi-IMPT and ARC plan. We observed that the conformity index was similar for the two methods (0.76 for ARC and 0.78 for multi-IMPT). The max dose value, $D_{max}$, decreased from 110.74% for ARC to 106.6% for multi-IMPT. Furthermore, multi-IMPT achieved substantially lower BED to OAR than proton ARC. Notably, $BED_{50}$ to bladder decreased from 44.6 Gy (ARC) to 24.49 Gy (multi-IMPT), and the $BED_{50}$ to rectum dropped nearly 45% from 45.02 Gy (ARC) to 24.69 Gy (multi-IMPT). Comparison of dose plots and DVH plots in Figure 1 indicates a slightly better dose distributions for target as well as OAR using multi-IMPT. Thus, for the prostate case, multi-IMPT plan provided a slightly improved overall performance.

**Lung:** Table 2 provides the lung case results. Comparing the results for both models, we can summarize that both plans have similar results. While the BED delivered to the OAR was slightly better for ARC (for example, $BED_{mean}$ = 6.24 Gy (ARC), 6.97 Gy (multi-IMPT), $BED_{mean}$ = 2.28 Gy (ARC), 2.54 Gy (multi-IMPT)), we observed that $D_{max}$ values slightly improved 113.39 from ARC to 109.31% for the multi-IMPT plan. However, the difference between the performance of the plans was not significant, thus, leading to nearly equivalent dose plans in this case. The plan equivalency is also evident from the dose plots and DVH plots given in Figure 2.

**Brain:** From Table 3, we observed that, in the brain case, ARC model slightly outperformed the multi-IMPT model in terms of the BED delivered to the OAR and CI (0.904 for ARC and 0.863 for multi-IMPT). Figure 3 also shows that the ARC model was slightly better than the multi-IMPT model in terms of DVH in OAR and target. However, the dose plots in Figure 3 show similar dose distribution for the two methods. Thus, while ARC outperforms multi-IMPT in the brain case, the difference was not significant.

**HN:** Table 4 shows the comparison of the two models for HN case. From the table, we observed that there is a <1% difference in $D_{max}$ values for both values. Furthermore, we note that the $BED_{max}$ values for right parotid and oral cavity for both models differ by less than 0.5 Gy. We also note that the difference between $BED_{mean}$ value of oral cavity for both models is around 1.1 Gy. Finally, we note that the $BED_{mean}$ value increases by around 7% for oropharynx for multi-IMPT model. From the dose plots, it is evident that the dose distribution for both models is quite similar. Thus, we can state that, for the HN case, the two models provide fairly equivalent dose plans.

Table 1: Comparison of (a) ARC, and (b) our model multi-IMPT for the prostate case.

| Structure | Quantity | Upper bound on BED (Gy) | ARC | multi-IMPT |
|---|---|---|---|---|
| CTV | CI | - | 0.76 | 0.783 |
|  | $D_{max}$ | - | 110.74% | 106.6% |
| Bladder | $BED_{50}$ (Gy) | 40 | 44.60 | 24.49 |
|  | $BED_{20}$ (Gy) | 63 | 67.75 | 58.65 |
| Rectum | $BED_{50}$ (Gy) | 40 | 45.02 | 24.69 |
|  | $BED_{20}$ (Gy) | 63 | 65.70 | 52.46 |
|  | $BED_{10}$ (Gy) | 90 | 81.48 | 73.39 |
| Femoral head | $BED_{10}$ (Gy) | 90 | 2.73 | 4.14 |
| Penile bulb | $BED_{50}$ (Gy) | 40 | 12.85 | 11.51 |

Table 2: Comparison of (a) ARC, and (b) our model multi-IMPT for the lung case.

| Structure | Quantity | Upper bound on BED (Gy) | ARC | multi-IMPT |
|---|---|---|---|---|
| CTV | CI | - | 0.937 | 0.929 |
|  | $D_{max}$ | - | 113.39% | 109.31% |
| Lung | $BED_{mean}$ (Gy) | 25.2 | 6.24 | 6.97 |
|  | $BED_{30}$ (Gy) | 15.6 | 3.23 | 3.43 |
| Heart | $BED_{mean}$ (Gy) | - | 2.28 | 2.54 |
| Esophagus | $BED_{mean}$ (Gy) | 28.66 | 5.06 | 5.29 |

Table 3: Comparison of (a) ARC, and (b) our model multi-IMPT for the brain case.

| Structure | Quantity | Upper bound on BED (Gy) | ARC | multi-IMPT |
|---|---|---|---|---|
| CTV | CI | - | 0.904 | 0.863 |
|  | $D_{max}$ | - | 103.14% | 104.08% |
| Brainstem | $BED_{max}$ (Gy) | 83.7 | 85.25 | 87.91 |
|  | $BED_{mean}$ (Gy) | - | 14.95 | 16.88 |
| Brain | $BED_{max}$ (Gy) | 96 | 104.31 | 109.37 |
|  | $BED_{mean}$ (Gy) | - | 1.01 | 1.11 |

Table 4: Comparison of (a) ARC, and (b) our model multi-IMPT for the HN case.

| Structure | Quantity | Upper bound on BED (Gy) | ARC | multi-IMPT |
|---|---|---|---|---|
| CTV | CI | - | 0.85 | 0.796 |
|  | $D_{max}$ | - | 102.19% | 102.45% |
| R Parotid | $BED_{max}$ (Gy) | - | 5.622 | 5.85 |
| Oral Cavity | $BED_{max}$ (Gy) | - | 140.5 | 140.21 |
|  | $BED_{mean}$ (Gy) | 66.85 | 10.95 | 12.14 |
| Oropharynx | $BED_{mean}$ (Gy) | 90.71 | 55.97 | 60.19 |
| Larynx | $BED_{mean}$ (Gy) | 78.42 | 4.73 | 5.13 |

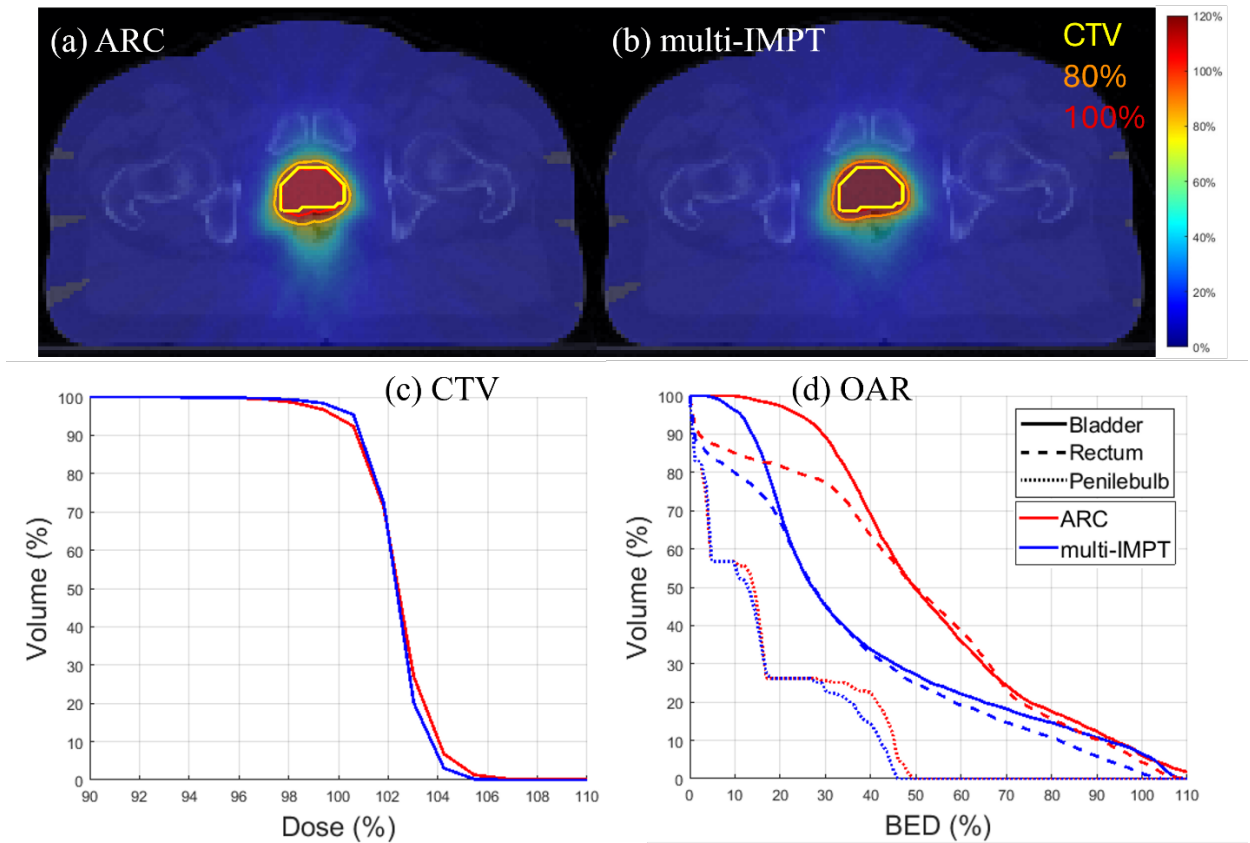

Figure 1. **Prostate**. (a), (b) Dose plots for ARC and multi-IMPT methods respectively, (c) DVH plot for the target, (d) BED-DVH plot for OAR

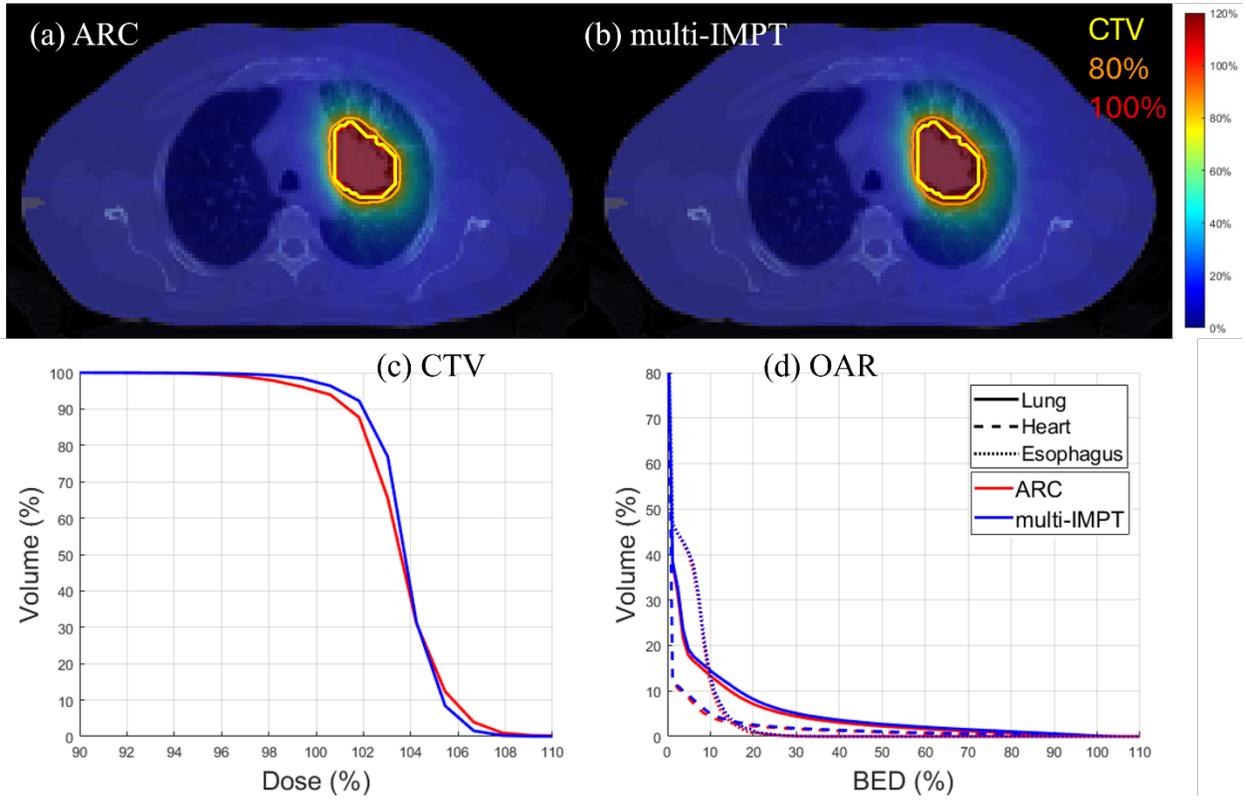

Figure 2. **Lung**. (a), (b) Dose plots for ARC and multi-IMPT methods respectively, (c) DVH plot for the target, (d) BED-DVH plot for OAR

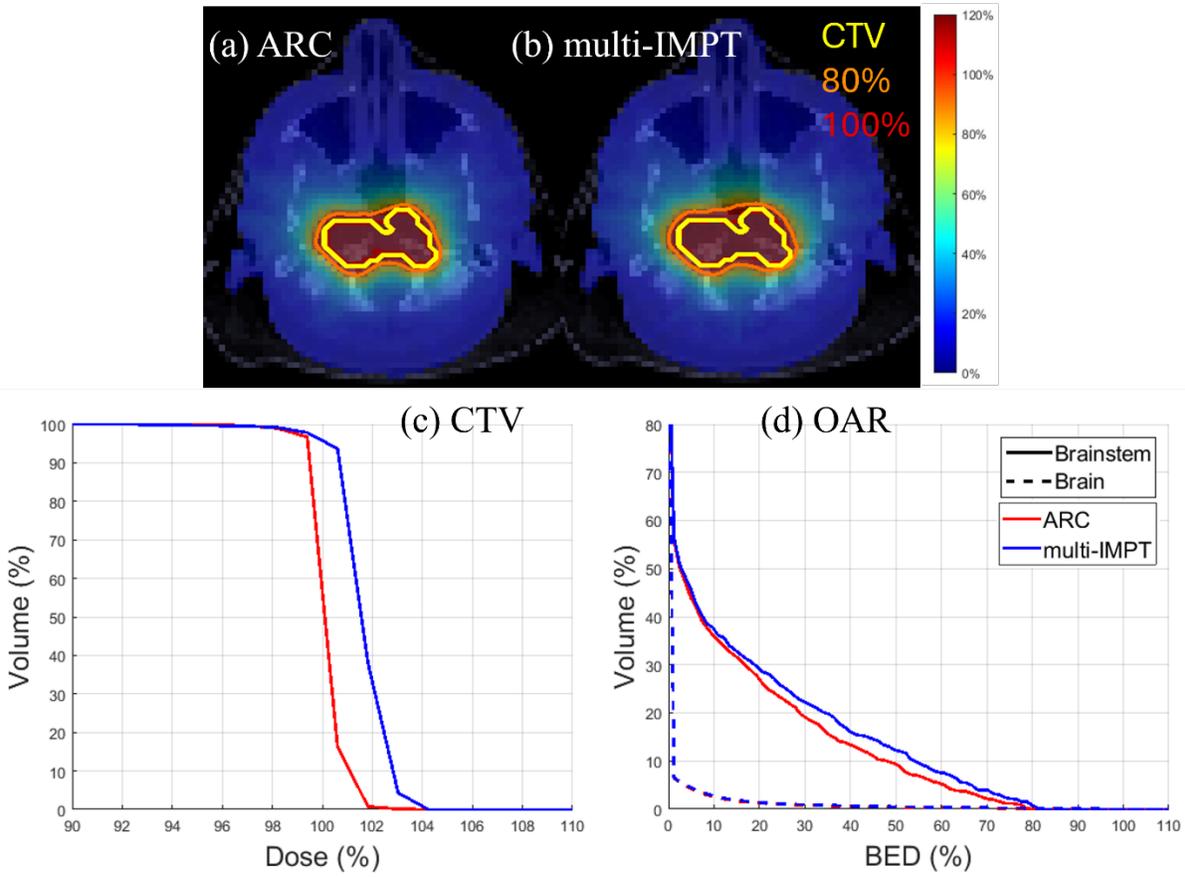

Figure 3. **Brain**. (a), (b) Dose plots for ARC and multi-IMPT method respectively, (c) DVH plot for the target, (d) BED-DVH plot for OAR

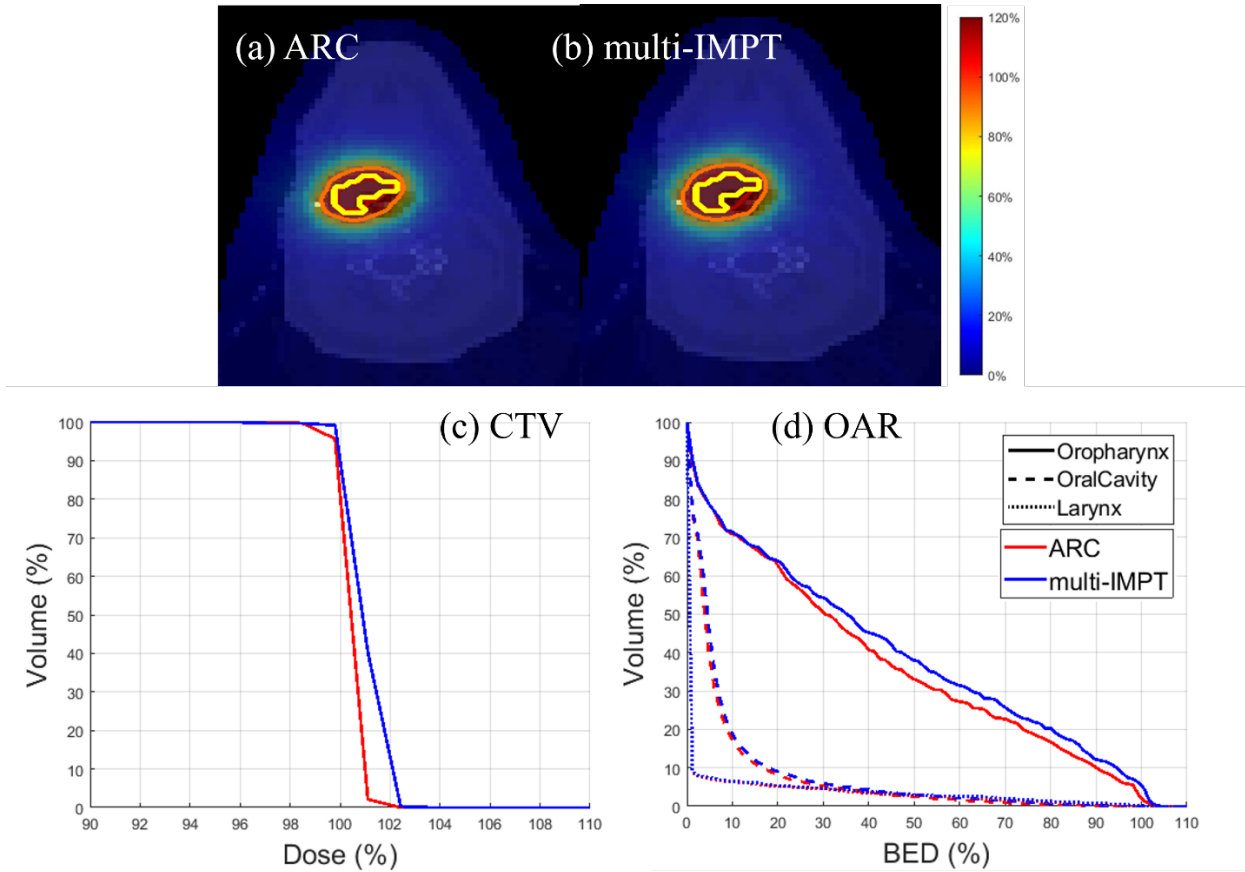

Figure 4. **HN**. (a), (b) Dose plots for ARC and multi-IMPT methods respectively, (c) DVH plot for the target, (d) BED-DVH plot for OAR

## 4. Conclusion and Discussion

In this work, we propose an ARC-equivalent IMPT method, termed multi-IMPT, which consists of multiple IMPT plans. The multi-IMPT method utilizes different combinations of a small subset of beams in each fraction, which sums up to a large set of beams as in ARC. It was shown that the multi-IMPT can deliver the dose coverage equivalent to ARC in terms of the BED delivered to the OAR and the physical dose delivered to the target.

From the results of four clinical cases, we observed that the BED delivered to the OAR in multi-IMPT was similar to the BED delivered to the OAR in proton ARC for two cases (brain and lung). Moreover, the BED delivered to the OAR in multi-IMPT was much lower compared to that of ARC plan for the prostate case. The physical dose delivered to the target using multi-IMPT plan matched the dose delivered using ARC for prostate and lung case. The DVH plot for the HN case showed an improvement in the physical dose delivery to the target. Thus, for three out of four cases, we observed that multi-IMPT was equivalent to or slightly better than ARC. For the brain case, multi-IMPT provided a slightly worse performance than ARC. Overall, the dosimetric difference multi-IMPT and ARC is not clinically significant, which shows that multi-IMPT can provide equivalent plan quality to ARC.

In multi-IMPT, the choice of each individual plan and the number of times a plan is used during the treatment impacts the BED delivered to the OAR. In this work, we do not compare the performance of individual dose plans generated by multi-IMPT. It might be possible to consider an optimal choice of plans for each fraction to minimize the overall BED delivered to the OAR.

**Appendix A: ICR and ADMM method (Algorithm 1) for solving the Augmented Lagrangian formulation (Eq. (3))**

In this section, we review the steps of the ADMM method (Algorithm 1) in detail. In Step 1, we carefully choose the values of the parameters in the augmented Lagrangian model. In Step 2, we can randomly initialize the decision variable $u_t$. We choose to set $u_t = 0$ $\forall t$. In Step 3, we initialize the remaining primal and dual variables. The Step 4 of the algorithm is performed $K$ times, i.e., we run the ADMM method for $K$ iterations.

Next, in each iteration $k$, Step 4a defines the active index set of the BED-DVH max and BED-DVH min constraints as explained in Section 2.1. The BED-DVH and DVH-min constraints are defined for OAR voxels and target voxels respectively based on the active index set. Finally, Step 4c updates the dual variables in the augmented Lagrangian formulation (Eq. (3)) and Step 4b defines the updates to the primal variables. We now describe the procedure to update each primal variable in detail.

1. Updating $u_t$: For each $t \in [T]$, we fix all the variables except $u_t$. The augmented Lagrangian formulation (Eq. (3)) is then unconstrained in $u_t$. Thus, we take the first-order derivative of the objective function in Eq. (3) and set it to 0. Then, $u_t$ is the solution of the resulting linear system of equations.

2. Updating $y_t$: For each $t \in [T]$, we fix all the variables except $y_t$ in Eq. (3). The resulting optimization problem has a closed form solution as follows: $y_t = max(g, u_t + \zeta_t)$ if $u_t + \zeta_t \geq g/2$. Otherwise, $y_t = 0$.

3. Updating $z^{tm}$ for all $m \in M_1$: For each $m \in M_1, j \in [n_m]$, we fix all the variables except $z_j^{tm}$ in Eq. (3). The resulting minimization problem over $z_j^{tm}$ is

$$\min \sum_{t=1}^{T} \left(A_j^{tm} u_t + \lambda_j^{tm} - z_j^{tm}\right)^2$$

$$\text{s.t.} \sum_{t=1}^{T} \left(z_j^{tm} + \frac{1}{2\rho_m}\right)^2 \leq \frac{BED_{max}^m}{\rho_m} + \frac{T}{4\rho_m^2}.$$

The optimal solution to the resulting minimization problem is the projection of $A_j^{tm} u_t + \lambda_j^{tm}$ onto the quadratic inequality constraint. The projected point (i.e., the optimal solution) is defined as $z_j^{tm} = (1-r)\left(A_j^{tm} u_t + \lambda_j^{tm}\right) - r\frac{1}{2\rho_m}$, where $r = max\{0, q\}$ and $q$ can be defined as follows:

$$q = 1 - \sqrt{\frac{\frac{BED_{max}^m}{\rho_m} + \frac{T}{4\rho_m^2}}{\sum_{t=1}^{T}\left(A_j^{tm} u_t + \lambda_j^{tm} + \frac{1}{2\rho_m}\right)^2}}.$$

4. Updating $z^{tm}$ for all $m \in M_2$: For each $m \in M_2$, we fix all the variables except $z_j^{tm}$ in Eq. (3). The resulting minimization problem over $z_j^{tm}$ is

$$\min \sum_{t=1}^{T} \sum_{j=1}^{n_m} \left(A_j^{tm} u_t + \lambda_j^{tm} - z_j^{tm}\right)^2$$

$$\text{s.t.} \sum_{t=1}^{T} \sum_{j=1}^{n_m} \left(z_j^{tm} + \frac{1}{2\rho_m}\right)^2 \leq \frac{n_m BED_{mean}^m}{\rho_m} + \frac{T n_m}{4\rho_m^2}.$$

We observe that the optimal solution to this problem is the projection of $A_j^{tm} u_t + \lambda_j^{tm}$ onto the inequality constraint. Thus, the optimal solution is $z_j^{tm} = (1-r)\left(A_j^{tm} u_t + \lambda_j^{tm}\right) - r\frac{1}{2\rho_m}$, where

$$r = \max\{0, q\} \text{ and } q = 1 - \sqrt{\frac{\frac{n_m BED_{mean}^m}{\rho_m} + \frac{T n_m}{4\rho_m^2}}{\sum_{t=1}^{T} \sum_{j=1}^{n_m} \left(A_j^{tm} u_t + \lambda_j^{tm} + \frac{1}{2\rho_m}\right)^2}}.$$

5. Updating $z^{tm}$ for all $m \in M_3$: For each $m \in M_1, j \in \Omega^m$, we use the same procedure as outlined for updating $z^{tm}$ for $m \in M_1$.

6. Updating $z^{t0}$: For each $j \in [n_0]$, we fix all variables except $z_j^{t0}$ in Eq. (3). The resulting minimization problem is

$$\min \sum_{t=1}^{T} \left(A_j^{t0} u_t + \lambda_j^{t0} - z_j^{t0}\right)^2$$

$$\text{s.t.} \sum_{t=1}^{T} \left(z_j^{t0} + \frac{1}{2\rho_0}\right)^2 \geq \frac{BED_{dv}^0}{\rho_0} + \frac{T}{4\rho_0^2}.$$

The optimal solution to this problem is the projection of $A_j^{t0} u_t + \lambda_j^{t0}$ onto the constraint. Thus, we can write the optimal solution to the above problem as $z_j^{t0} = (1+r)\left(A_j^{t0} u_t + \lambda_j^{t0}\right) + r\frac{1}{2\rho_0}$, where

$$r = \max\{0, q\} \text{ and } q = \sqrt{\frac{\frac{BED_{dv}^0}{\rho_0} + \frac{T}{4\rho_0^2}}{\sum_{t=1}^{T} \left(A_j^{t0} u_t + \lambda_j^{t0} + \frac{1}{2\rho_0}\right)^2}} - 1.$$